\documentclass[runningheads]{llncs}
\usepackage{graphicx}
\usepackage{kotex}

\usepackage{tabu}
\usepackage{amsmath}

\usepackage{color}

\begin{document}
\title{Anomaly Detection for Industrial Control Systems Using Sequence-to-Sequence Neural Networks}

%
\titlerunning{Anomaly Detection for ICSs Using Seq2Seq Neural Networks}

\author{Jonguk Kim \and Jeong-Han Yun \and Hyoung Chun Kim}

%

\authorrunning{J. Kim et al.}

%
\institute{The Affiliated Institute of ETRI, Republic of Korea 
\email{\{jongukim,dolgam,khche\}@nsr.re.kr}}

\maketitle              
\begin{abstract}
This study proposes an anomaly detection method for operational data of industrial control systems (ICSs).
Sequence-to-sequence neural networks were applied to train and predict ICS operational data and interpret their time-series characteristic.
The proposed method requires only a normal dataset to understand ICS's \textit{normal} state and detect outliers.
This method was evaluated with SWaT (secure water treatment) dataset, and 29 out of 36 attacks were detected.
The reported method also detects the attack points, and 25 out of 53 points were detected.
This study provides a detailed analysis of false positives and false negatives of the experimental results.

\end{abstract}

\section{Introduction}

Since Stuxnet struck a nuclear facility in 2010, threats towards industrial control systems (ICSs) have increased. Unfortunately, as most of ICS attackers are state-sponsored and use zero-day vulnerabilities, signature-based detection (maintaining the blacklist and updating it) is inappropriate.

The most common and safe approaches that do not harm the availability of ICSs monitor the network traffic of these systems.
ICSs present more periodic behavior than information technology systems.
Several studies \cite{formby2016s,lin2017timing,yun2018similarity} applied the statistical characteristics of traffic for ICS-specific security mechanisms. 
Although this approach is suitable for ICS traffic characteristics,
    it presents limitations in detecting attacks at the ICS operation level.

Other researches focused on the detection of anomalies with physical properties \cite{mitchell2014survey}. 
By using the specification or control logic, the monitoring system rarely emits false alarms~\cite{chen2018learning,mitchell2013behavior}. 
However, it is relatively expensive to obtain and specify the specification or control logic.
An ICS recognizes the environment with sensors, makes decisions for its purpose, and delivers the right action with actuators.
To detect anomalies at the ICS operation level, its \textit{normal} state must be defined and the control logic that decides actuators' behaviors must be understood.
However, understanding the entire set of the control logic is complicated.
In fact, the volume of the control logic is enormous, and acquiring it from vendors is not allowed in most cases.

Herein, the aim is to monitor the ICS operational data.
A feasible alternative is the \textit{data-driven} approach.
Machine-learning-based approaches have been highly studied and especially deep-learning-based anomaly detection methods which have been reported recently using fully-connected networks (FCN) \cite{kaspersky}, convolutional neural networks (CNN) \cite{israel}, recurrent neural networks (RNN) \cite{goh2017anomaly}, and generative adversarial networks (GAN) \cite{li2019mad}.

We propose a deep learning-based anomaly detection method using a sequence-to-sequence model (seq2seq) \cite{seq2seq}.
Seq2seq is designed initially for natural language translation.
It encodes the words of a given sentence with RNN into a latent vector, then decodes from it to a set of words in the target language.
Seq2seq's encoding-decoding approach presents a significant advantage as it can understand the context of the entire sentence, while vanilla RNN gives the output immediately for every input.
Seq2seq is expected to be an effective method for learning the context or semantics of time-series operational data, and obtain a better prediction based on the given data.

To date, no abnormal samples have been reported to train machine learning models robustly.
Therefore, the reported model is trained with the only normal dataset (training dataset), and it is considered that the training data are clean.
In the detection phase, the developed model investigates unseen data with trained neural networks.
Using the model after the learning phase, the detection method proceeds in three steps: Step 1, the model predicts the future values of the sensors and actuators, Step 2, the difference between the prediction and actual data is determined, and Step 3, alerts are sent for significant differences.

The rest of this paper is organized as follows.
Section~\ref{sec:method} introduces the anomaly detection method using the seq2seq neural network.
Section~\ref{sec:experiment} presents the experimental results after applying the proposed method to the secure water treatment (SWaT) dataset~\cite{goh2016dataset}.
Section~\ref{sec:analysis} analyzes the experimental results in detail and Section~\ref{sec:conclusion} concludes this study.

\section{Proposed Method}
\label{sec:method}

\subsection{SWaT Dataset}
\label{sec:method-swat-dataset}

Several studies have recently been reported on dataset generation for ICS research~\cite{choi2018dataset,goh2016dataset,lemay2016providing,rodofile2017process}. 
The most frequently used dataset is the SWaT dataset~\cite{goh2016dataset} by Singapore University of Technology and Design (SUTD), which has operational data and attack labels.
Herein, the method is developed and evaluated with the SWaT dataset.

The SWaT dataset was collected from a testbed water treatment system.
Fifty-one tags (25 sensors and 26 actuators) are sampled every second.
Some are digital, and  others are analog.
Tag names define their roles.
For example, MV denotes motorized valve, P for pump, FIT for flow meter, and LIT for level transmitter.

SWaT consists of six processes.
Water flows from process 1 to the process 6.
The numbers following the tag names indicate the process ID and the gadget ID.
For example, MV-101 is the first motorized valve in process 1.

In the SWaT dataset, normal and attack datasets are separated.
The normal part has 480,800 samples, 
    and the attack part includes 41 attacks during 449,919 samples. 

\subsection{Data Preprocessing}
\label{sec:preprocessing}

Multiple machine learning schemes normalize the input into a Gaussian distribution with an average of 0 and a standard deviation of 1.
However, to ensure that the distributions of most tags in the normal part of the SWaT dataset were not distorted nor had multiple peaks, min-max normalization was chosen.
Long-short term memory (LSTM)~\cite{lstm}, which is a RNN cell used in the developed model, has a sigmoid function inside that gives a (0, 1) output.
The minimum and maximum for normalization were chosen as 0 and 1.

The operational data were time-series.
The model used sliding windows of length 100 seconds to understand the temporal context, and each window slides 1 by 1 second.

\subsection{Sequence-to-Sequence Neural Networks}
\label{sec:method-model}

\begin{figure}[t]
    \centering
    \includegraphics[width=\textwidth]{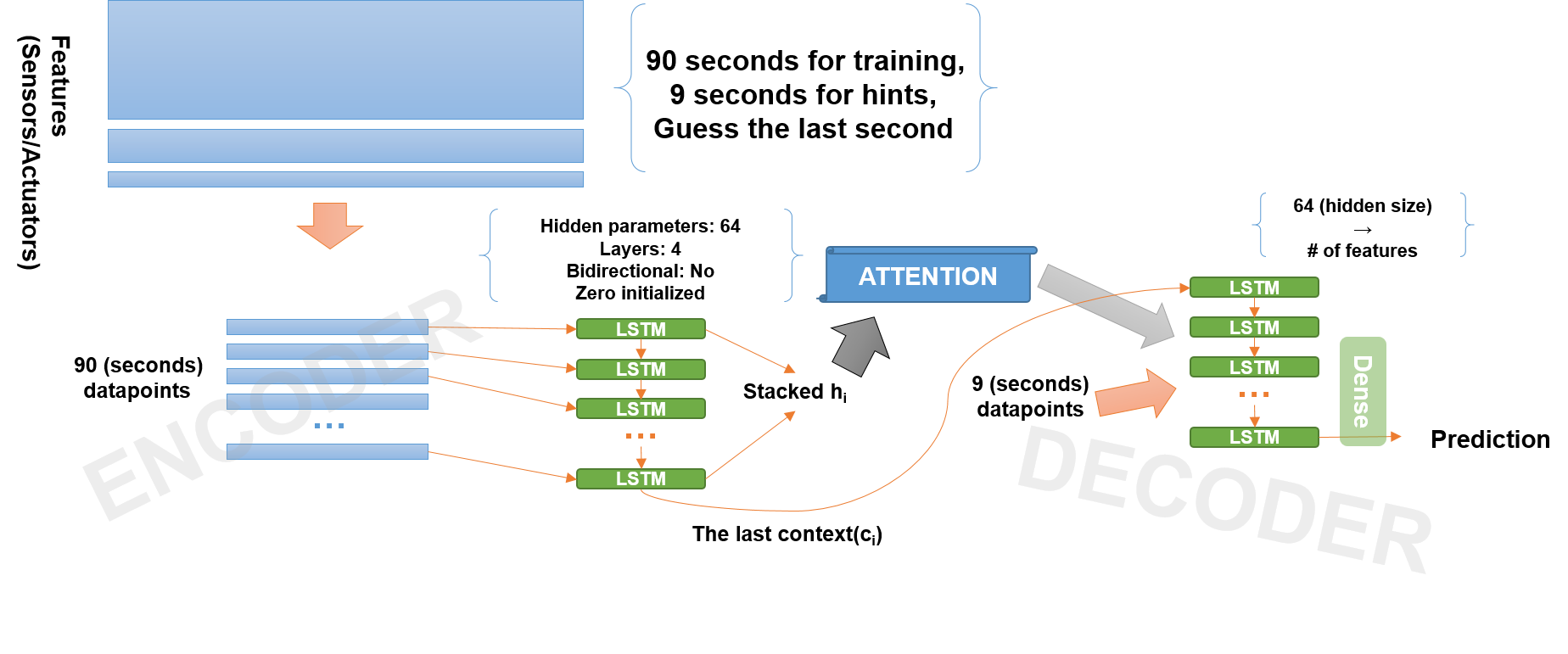}
    \caption{Proposed learning model using seq2seq with attention}
    \label{fig:model-arch}
\end{figure}

There are various ways to build a neural network for learning time-series data.
We chose the sequence-to-sequence network with attention~\cite{attention} for training and evaluation.
Since RNN is proper for learning a series of data, we expected that RNN is able to encode the current window of data and to anticipate the next values.
Figure~\ref{fig:model-arch} shows the data flow of the proposed model.
The shape of inputs for the encoder is (\# of sequences, batch, \# of tags).
The encoder obtains the first 90 seconds of the window.
The encoder's output is two-fold: 1) the last cell state of the last layer and 2) all hidden states of the last layer.
The first one has the context of the given sequences.
The second one helps the decoder predict future operational data with the attention mechanism.
The decoder part is optional.
The reason why we added a decoder with attention is that it gives us more accurate results.
The decoder predicts the last second with a 9-second hint.
The values of 9th second of the hint is almost the same with those of the last record of the window (we wanted to expect), which helps the model give almost-zero prediction error.
The shape of the decoder's final output is (batch, \# of tags).

\begin{figure}[t]
    \centering
    \includegraphics[width=\textwidth]{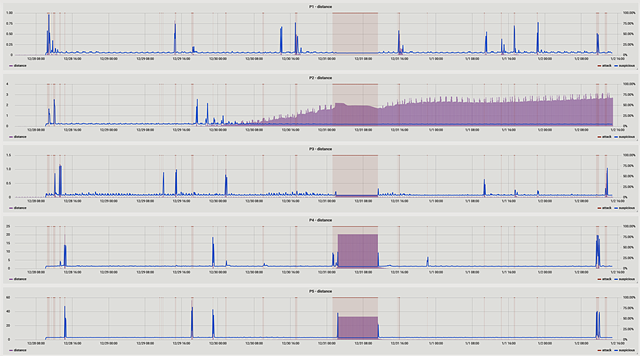}
    \caption{Prediction errors for processes 1, 2, 3, 4, and 5 of SWaT (red: attack, purple: prediction error, and blue: anomaly score)}
    \label{fig:whole-shot}
\end{figure}

An independent model was applied for each process of SWaT.
The model $n$ learns the process $n$.
Figure~\ref{fig:whole-shot} shows the prediction error of each process.
At the early stage of this study, a holistic model\footnote{\cite{kaspersky} used this approach: one model for the whole processes.} was tested for six processes.
The result was not accurate because each process shows a different prediction error pattern, especially the process 2, as observed in Figure~\ref{fig:whole-shot}.

\subsection{Measuring Prediction Error}
\label{sec:method-prediction-error}

The mean absolute error (MAE) is a typical measuring method.
However, $p$-norm can also be used, with $p >$ 2.
The greater the value of $p$, the greater the value of the vector.
$\infty$-norm returns the maximum value of the vector.
Herein, a 4-norm was chosen, while Kaspersky chose a 6-norm~\cite{kaspersky} for SWaT dataset.

The prediction error (distance) $D$ can be extracted as follows:
\begin{equation}
    D = \sqrt[4]{\sum_{i=1}^{n} d_{i}^{4}} \text{ where } d_{i} = I_{i} - O_{i}
    \label{eq:norm}
\end{equation}
where $n$ is the number of tags, $I_{i}$ is the $i$-th tag values in the dataset, and $O_{i}$ is the $i$-th output of the model.

\subsection{Anomaly Decision using Prediction Errors}
\label{sec:method-decision}

The proposed method considers that the system is under attack if the model has never seen the current state.
The developed model was trained to perform a precise prediction.
When the model detects a never-seen window, it cannot perform an accurate prediction, which leads to more notable $4$-norm value.

Multiple approaches can be used to determine anomalies such as cumulative sum (CUSUM) and anomaly likelihood \cite{ahmad2017anomaly}.
The custom rating method was applied by considering the prediction errors -- due to the following factors:
\begin{enumerate}
    \item A learning model often presents periodic noises, but it is hard to remove noises because of the limited learning data.
    \item From a specific time, the distances of process 2, especially AIT-201 in SWaT, are increasing and never recovered (this issue is discussed in Section \ref{sec:analysis-false-positives}).
\end{enumerate}

A similar (but not significant) phenomenon occurred in process 5.
It is assumed that an insufficient amount of training data and unexpected dynamics can cause growing prediction errors.

A sliding window for rating was also applied.
First, the top-$k$ outliers were removed.
As aforementioned, the developed model presents noise (the impulse of prediction error).
The sum of remains except the outliers represents the distance of the window.
Fortunately, attacks last at least 2 minutes (the shortest attack is attack 34).
For a large summation on a specific window, a high rate can be attributed to that particular region.

Second, $N$ summations were collected to compare the current sum with near-by ones.
Two hyper-parameters were defined: $H$ and $L$.
$L$ is the 20th percentile, and $H$ is the 90th percentile.
$H$ is not the 80th because top-k values (outliers) were already removed to reduce noises.
The ratio $\frac{H}{L}$ and divided by $R$, which is another hyper-parameter.
Here, $R = $ 20.

In short, if the 90th percentile is 20 times greater than the 20th percentile, it is certain that an attack happens.
The rate for suspicious region $S$ is derived as follows:
\begin{equation}
    S = min(\frac{H}{LR}, 1.0)
    \label{eq:rating}
\end{equation}

Finally, if $S$ is greater than 0.3 (30\%), it is regarded as an attack.
All numbers mentioned above (90 for $H$, 20 for $L$, 20 for $R$, and 0.3 for threshold) are hyper-parameters which are dataset-dependent.

\begin{figure}
    \centering
    \includegraphics[width=\textwidth]{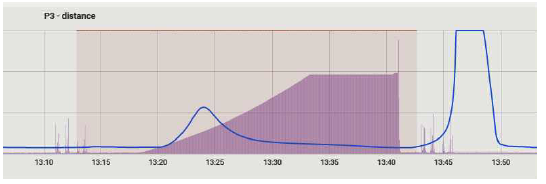}
    \caption{Asymmetric ratings at the start and end of attack 41 (red: attack, purple: prediction error, and blue: anomaly score)}
    \label{fig:attack41}
\end{figure}

Equation~\ref{eq:rating} determines high rates at the start and end of the attack because it measures the change.
We use both sides: start and end.
Depending on the attack the high rate can be obtained at the start or at the end.
Figure~\ref{fig:attack41} shows $D$ and $S$ of attack 41.

As $D$ grows slowly at the start and shrink quickly at the end, $S$ is low at the start but high at the end.

\section{Experiment}
\label{sec:experiment}

All source codes, pre-processed datasets, trained network parameters, and results are available at \url{https://github.com/jukworks/swat-seq2seq}.

\subsection{Training}

Occasionally, a neural network goes bad local minima during training process.
A general approach to solve this issue is to train the neural network multiple times independently and choose the best result among the multiple training results.
As a neural network is trained with a stochastic gradient descent and mostly initialized with random numbers, different results are obtained every run.
Two independent training sessions were run for each model (each process) and the network presenting the lowest training loss was chosen.

The model was optimized with Adam~\cite{adam}, amsgrad~\cite{amsgrad}, and without weight decay.
Each model trained 150 epochs with a 4,096 batch size.
Early stopping was not applied.

\begin{table}[t]
\caption{Training time for 150 epochs}
\begin{center}
\begin{tabu}{X[c]|X[c]|X[c]|X[c]|X[c]|X[c]|X[c]}
\hline
 & Process 1 & Process 2 & Process 3 & Process 4 & Process 5 & Process 6 \\
\hline\hline
Trial 1 & 1h 48m & 2h 00m & 1h 53m & 1h 54m & 2h 23m & 2h 27m \\
Trial 2 & 1h 50m & 1h 59m & 1h 59m & 2h 01m & 2h 01m & 2h 12m \\
\hline
\end{tabu}
\end{center}
\label{tab:training-time}
\end{table}

The hardware consisted of Intel Xeon CPU E5-2960 v4 2.60GHz, 6 NVIDIA Tesla V100, and 512 GB RAM.
Table~\ref{tab:training-time} shows that the training time was approximately 2 hours on average.
Six models have a different size of input and output, but their internal LSTM architectures were entirely the same.
Therefore, the number of trainable parameters is similar, which leads to similar training time.

\subsection{Anomaly Detection}
The results are compared with Kaspersky's research~\cite{kaspersky} because this is the only study providing the list of found attack points.

\begin{figure}[t]
    \centering
    \includegraphics[width=0.6\textwidth]{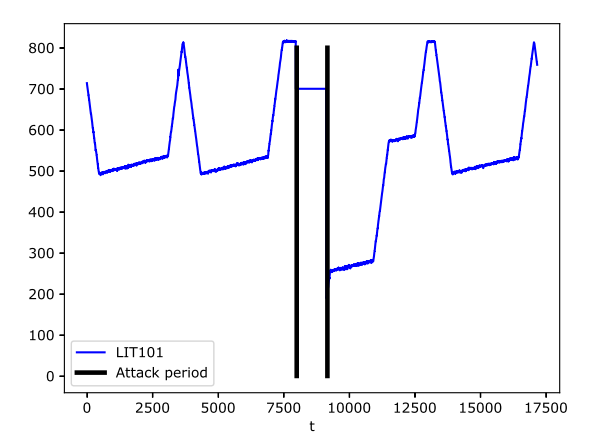}
    \caption{LIT-101 was not stable after attack 30 happens (from \cite{israel})}
    \label{fig:Israel-attack30}
\end{figure}

If an alert is received within 15 minutes after the attack range of SWaT, it is considered as a detection.
15 minute is chosen as attacks attacks on cyber-physical systems tend to have a long impact.
The shortest attack in SWaT, attack 29, lasts 2 minutes\footnote{The longest attack in SWaT is attack 28, which lasted 9.5 hours.}.
Figure~\ref{fig:Israel-attack30} shows LIT-101 on attack 30~\cite{israel}.
The two black vertical lines represent the start and end of the attack.
After the end of the attack, LIT-101 was unstable.
As the reported model learns the normal-labeled data only, it may perceive the stabilizing region as an anomaly.

Tables~\ref{tab:result-table-1} and \ref{tab:result-table-2} compare the detection results with those of~\cite{kaspersky}.
Attacks 5, 9, 12, 15, and 18 have no physical impact.
These attacks were ignored as they cannot be detected with operational data.
The first column shows the attack numbers.
The second column presents the answer to attack points labeled in the SWaT dataset.
The third column represents the detection of the attack: \textit{Yes} means 100\% sure, 
\textit{not sure} means 30\% - 100\% means sure, and \textit{No} means less than 30\% sure.
As mentioned in Section \ref{sec:method-decision}, 30\% represents the heuristic threshold.
The fourth and fifth column represents the attack points determined by the model and~\cite{kaspersky}, respectively.
Bold text indicates correct answers.
The parentheses indicate the second-longest distance.
\textit{N/A} indicates that the model failed to detect.

\begin{table}
\caption{Anomaly detection results compared with those of~\cite{kaspersky} (attacks 1 to 30)}
\begin{center}
\begin{tabu} to 0.9\textwidth { X[c] | X[c] | X[c] | X[c] | X[c] }
\hline
Attack \# & Answer (by SWaT) & Detection (our work) & Attack point (our work)  & Attack point (\cite{kaspersky}) \\
\hline\hline
1 & MV-101 & Yes & \textbf{MV-101} & N/A \\
\hline
2 & P-102 & Yes & MV-101 (\textbf{P-102}) & MV-301 (\textbf{P-102}) \\
\hline
3 & LIT-101 & Not sure (65\%) & MV-101 (\textbf{LIT-101}) & N/A \\
\hline
4 & MV-504 & No & N/A & N/A \\
\hline
6 & AIT-202 & Yes & \textbf{AIT-202} & \textbf{AIT-202} (P-203) \\
\hline
7 & LIT-301 & Yes & \textbf{LIT-301} & \textbf{LIT-301} (PIT-502) \\
\hline
8 & DPIT-301 & Yes & \textbf{DPIT-301} & \textbf{DPIT-301} (MV-302) \\
\hline
10 & FIT-401 & Yes & \textbf{FIT-401} & \textbf{FIT-401} (PIT-502) \\
\hline
11 & FIT-401 & Yes & \textbf{FIT-401} (FIT-504) & MV-304 (MV-302) \\
\hline
13 & MV-304 & No & MV-304 & N/A \\
\hline
14 & MV-303 & No & N/A & N/A \\
\hline
16 & LIT-301 & Yes & \textbf{LIT-301} & MV-301 (MV-303) \\
\hline
17 & MV-303 & Yes & MV-301 (\textbf{MV-303}) & N/A \\
\hline
19 & AIT-504 & No (15\%) & AIT-504 & \textbf{AIT-504} (P-501) \\
\hline
20 & AIT-504 & Yes & \textbf{AIT-504} & N/A \\
\hline
21 & MV-101, LIT-101 & Not sure (35\%) & \textbf{LIT-101} & UV-401 (P-501) \\
\hline
22 & UV-401, AIT-502, P-501 & Yes & FIT-401, FIT-504 & DPIT-301 (MV-302) \\
\hline
23 & P-602, DPIT-301, MV-302 & Yes & \textbf{DPIT-301} & P-302, P-203 \\
\hline
24 & P-203, P-205 & No & N/A & LIT-401 \\
\hline
25 & LIT-401, P-401 & No (20\%) & LIT-401 & P-602, MV-303 \\
\hline
26 & P-101, LIT-301 & Yes ~~~~~~~ (25\% at P3) & P-102, LIT-301 & LIT-401 (AIT-402) \\
\hline
27 & P-302, LIT-401 & Yes & \textbf{LIT-401} & N/A \\
\hline
28 & P-302 & Yes & FIT-401, AIT-504 & MV-201, LIT-101 \\
\hline
29 & P-201, P-203, P-205 & No & N/A & LIT-401, AIT-503 \\
\hline
30 & LIT-101, P-101, MV-201 & Yes & \textbf{LIT-101} & LIT-301 (FIT-301) \\
\hline
\end{tabu}
\end{center}
\label{tab:result-table-1}
\end{table}

\begin{table}[t]
\caption{Anomaly detection results compared with those of~\cite{kaspersky} (attacks 31 to 41)}
\begin{center}
\begin{tabu} to 0.9\textwidth { X[c] | X[c] | X[c] | X[c] | X[c] }
\hline
Attack \# & Answer (by SWaT) & Detection (our work) & Attack point (our work)  & Attack point (\cite{kaspersky}) \\
\hline\hline
31 & LIT-401 & Not sure (50\%) & \textbf{LIT-101} & P-602, MV-303 \\
\hline
32 & LIT-301 & Yes & \textbf{LIT-301} & N/A \\
\hline
33 & LIT-101 & Yes & \textbf{LIT-101} & N/A \\
\hline
34 & P-101 & Yes & \textbf{P-101} & MV-201 (P-203) \\
\hline
35 & P-101, P-102 & Not sure ~~~ (20\% at P1, 45\% at P3) & P-101 & MV-201, MV-303 \\
\hline
36 & LIT-101 & Yes & \textbf{LIT-101} & \textbf{LIT-101}, AIT-503 \\
\hline
37 & P-501, FIT-502 & Yes & FIT-401, FIT-504 & FIT-504 (FIT-503) \\
\hline
38 & AIT-402, AIT-502 & Yes ~~~~~~~ (15\% at P5) & MV-101, \textbf{AIT-402}, AIT-502 & \textbf{AIT-502}, \textbf{AIT-402} \\
\hline
39 & FIT-401, AIT-502 & Yes & \textbf{FIT-401} & \textbf{FIT-401}, P-201 \\
\hline
40 & FIT-401 & Yes & \textbf{FIT-401}, FIT-504 & UV-401 (\textbf{FIT-401}) \\
\hline
41 & LIT-301 & Yes & \textbf{LIT-301} & N/A \\
\hline
\end{tabu}
\end{center}
\label{tab:result-table-2}
\end{table}

\cite{kaspersky} reported 25 attacks and 11 attack points (nine with the first predictions and 2 with the second predictions).
Herein, the model found 29 attacks and 25 attack points (22 with the first predictions and 3 with the second predictions).
21 attacks were detected by both the developed model and that in~\cite{kaspersky}.
Four attacks were detected only by the model in~\cite{kaspersky}: attacks 19, 24, 25, and 29.
Eight attacks were detected only by the developed model: attacks 1, 3, 17, 20, 27, 32, 33, and 41, which also detected attacks 21 and 31 with 35\% and 50\% rates, respectively.
The SWaT dataset indicates that the attack points of attack 35 is process 1, but it was detected by the developed model for process 3 (20\% rate at the model for process 1).
Both methods failed to detect three attacks: attacks 4, 13, and 14.

\begin{figure}[t]
    \centering
    \includegraphics[width=0.6\textwidth]{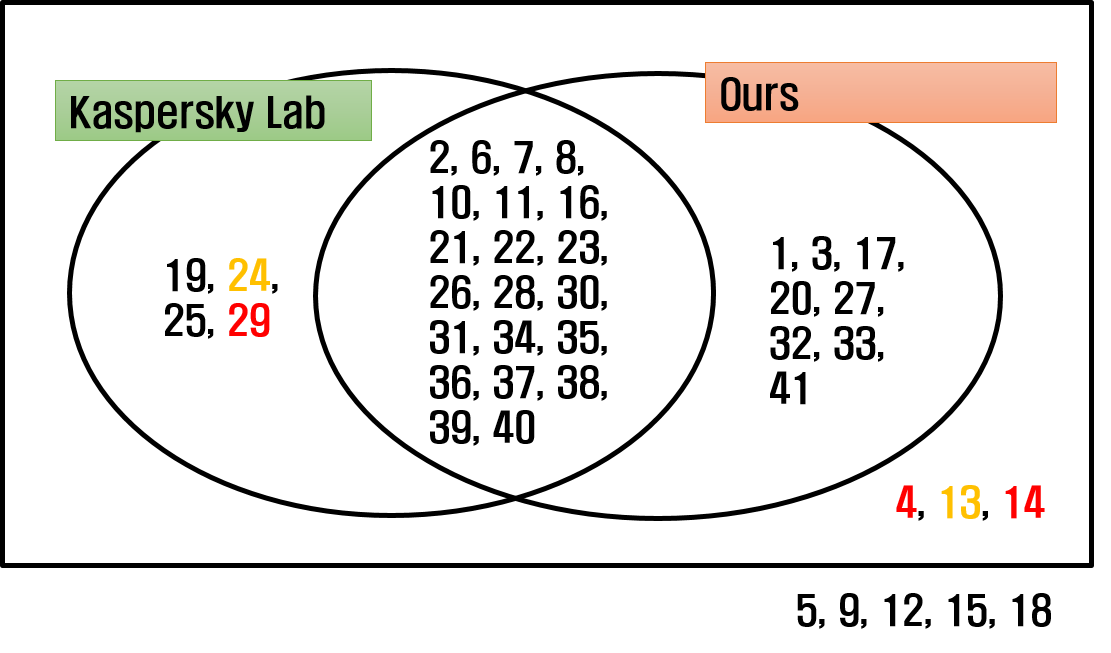}
    \caption{List of detected attacks and comparison with~\cite{kaspersky}. Red (4, 14, 29) represents attacks that are impossible to detect. Yellow (24) represents attacks that are difficult to detect.}
    \label{fig:venn}
\end{figure}

The model was not used with process 6 as it has only two tags.
The SWaT dataset presents only one attack\footnote{Attack 23 also has an impact on process 3. The developed model for process 3 detected this attack.} and  has an impact on process 6.

\section{Analysis of Experimental Results}
\label{sec:analysis}

Three pairs of attacks, (10, 11), (19, 20), and (27, 28) were indistinguishable for the developed model because they occur almost continuously\footnote{Their intervals are 0, 42, and 1 second(s) respectively, while the developed model uses a 100-second sliding windows.} and have the same attack point.

\subsection{Analysis on False Negative (Undetected Attacks)}
\label{sec:analysis-FN}

The model failed to detect attacks 4, 13, 14, 19, 24, 25, and 29.

\subsubsection{Impossible-to-find attacks}
It appears that attacks 4, 14, and 29 were impossible to detect (in red in Figure~\ref{fig:venn}).
The developed model, the model in~\cite{kaspersky}, and the model in~\cite{israel} all failed to detect attacks 4, 13, and 14.

\paragraph{Attack 4}
This attack opens MV-504 that does not exist in the dataset.
The description~\cite{goh2016dataset} in SWaT's \textit{list of attacks} indicates that this attack has no impact.

\paragraph{Attack 14}
According to SUTD, the attack 14 failed because tank 301 was already full~\cite{goh2016dataset}.

\paragraph{Attack 29}
SUTD said that P-201, P-203, and P-205 did not start because of mechanical interlocks.
In the dataset, nothing was changed around 2015/12/31 at 3:32:00 PM.

\subsubsection{Difficult-to-find attacks}

According to~\cite{goh2016dataset}, attacks 13 and 24 have a small impact (represented in yellow in Figure~\ref{fig:venn}).
The developed model did not detect these attacks.
The model also failed to detect attacks 19 and 25.

\paragraph{Attack 13}
This attack attempted to close MV-304 but MV-304 was closed later than when this attack occurred.
In the SWaT dataset, MV-304 did not change.

\begin{figure}
    \centering
    \includegraphics[width=0.9\textwidth]{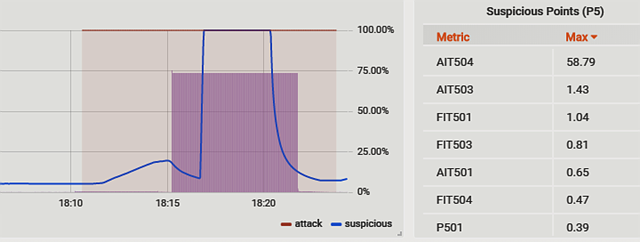}
    \caption{Attack 19 in process 5 (red: attack, purple: prediction error, and blue: anomaly score)}
    \label{fig:attack19}
\end{figure}

\paragraph{Attack 19}
This attack attempted to set value of AIT-504 to 16 $\mu$s/cm.
AIT-504 is below 15 $\mu$s/cm in the normal state.
In Figure~\ref{fig:attack19}, the attack appeared to be detected, but the high rate was derived from attack 20 which set a value of AIT-504 to 255 $\mu$s/cm.
the distances provided by the developed model are too short to detect for attack 19.
The rate $S$ was of approximately 20\% for attack~19.

\paragraph{Attack 24}
This attack attempted to turn off P-203 and P-205 (both are pumps are used for injecting chemicals).
However, there was only a small impact due to the closure of P-101.

\begin{figure}[t]
    \centering
    \includegraphics[width=\textwidth]{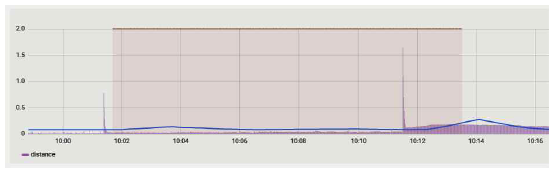}
    \caption{Attack 25 in process 4 (red: attack, purple: prediction error, and blue: anomaly score)}
    \label{fig:attack25_detail}
\end{figure}

\paragraph{Attack 25}
This attack attempted to set the value of LIT-401 to 1,000 and open P-402 while P-402 was still operating.
LIT-401 presented notable distances but the rate was of approximately 20\% (Figure~\ref{fig:attack25_detail}).
According to~\cite{kaspersky}, attack 25 was detected but wrong attack points were determined.

\subsubsection{Managed-to-find attack}

The developed model managed to detect attack 35 with a rate of 45\%.

\paragraph{Attack 35}
\begin{figure}
    \centering
    \includegraphics[width=\textwidth]{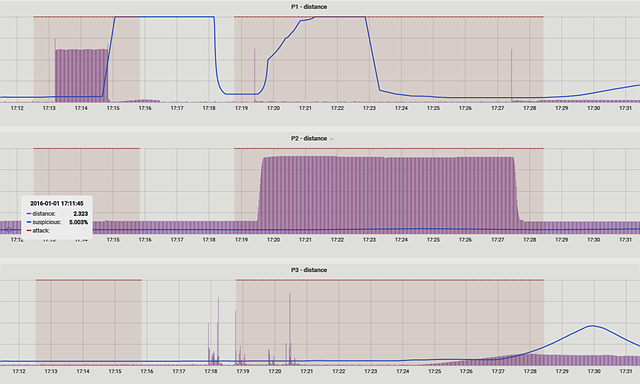}
    \caption{Attacks 34 and 35 in processes 1, 2, and 3 (red: attack, purple: prediction error, and blue: anomaly score)}
    \label{fig:attack34_35}
\end{figure}

According to~\cite{goh2016dataset}, attack 35 occurred at process 1, but herein, the attack was detected at process 3.
Before the attack, P-101 was open, and P-202 was close.
The attack opened P-101 and kept P-102 close.
Figure~\ref{fig:attack34_35} shows the attacks 34 and 35.
Processes 1, 2, and 3 presented remarkable distances.
The distance of process 2 appears to be large, but the scale is small.
There are two high peaks of rate in process 1, but they come from the attack 34.

Attack 34 closed P-101 ($2 \rightarrow 1$) and opened P-102 ($1 \rightarrow 2$).
Later, P-101 was opened ($1 \rightarrow 2$) at 17:14:59 and P-102 was closed ($2 \rightarrow 1$).
Attack 35 was different from attack 34 as it kept P-102 closed.
In the training (normal) dataset, P-102 was closed (of value 1) at all time, which is why the developed model always indicated 1.
Because P-102 is a backup pump for P-101, the developed model must understand their connection.
However, the model could not learn the connection as the training dataset did not give enough information.

\subsection{Analysis on False Positives (False Alarms)}
\label{sec:analysis-false-positives}

\begin{table}[t]
\caption{Number of false positives. OP indicates the number of false positives for attacks on other processes, LT indicates the long-tailed detection (over 15 minutes), and TFP indicates the true false positives.}
\begin{center}
\begin{tabu}{X[c]|X[c]||X[c]|X[c]|X[c]}
\hline
Process & False positives & OP & LT & TFP \\
~ & (all duplicates) & ~ & ~ & ~ \\
\hline\hline
1 & 7(11) & 4 & 2 & 1 \\
2 & 5(8) & 0 & 0 & 5 \\
3 & 0 & - & - & - \\
4 & 1(1) & 1 & 0 & 0 \\
5 & 0 & - & - & - \\
\hline
\end{tabu}
\end{center}
\label{tab:false_positives}
\end{table}

This subsection analyzes false positives. The evaluation script is also available at \url{https://github.com/jukworks/swat-seq2seq/tree/master/evaluation}.

Sixty seven false positives were obtained from all models, and many of them were timely overlapped.
After removing the duplicates, twenty false positives were found.
The number of false positives is presented in Table~\ref{tab:false_positives}.
Processes 3 and 5 present no false positives.

The decision method described in Section~\ref{sec:method-decision} reacted twice at the start and end of the attack.
The second column represents the number of false positives after removing the duplicates.
The third column, OP, represents the number of positives caused by the attacks on the other processes.
The fourth column, LT, represents the number of long-tailed positives.
15 minutes after the end of the attack as a true-positive, but sometimes the tail was too long and the attack lasted over 15 minutes.
The long-tail positives were not counted as true-positives.
In summary, there was one true false positive (TFP, the fifth column) in process 1 and five in process 2.

\subsubsection{False positives in process 1}

There were eleven false positives in process 1.
In Table~\ref{tab:false_positives_p1}, (2, 3), (4, 5), (7, 8), and (10, 11) are pairs of start-end having the same related attacks.
Therefore there were seven independent false positives.
One of them was true false positive.
Four of them were simultaneous detection for attacks targeting other processes. 
Two of them were long-tail detection.

\begin{table}[t]
\caption{The analysis of false positives in process 1. OP means the number of false positives for attacks on other processes; LT for long-tailed detection (over 15 minutes); TFP for true false positives.}
\begin{center}
\begin{tabular}{c|c|c|c|c}
\hline
No. & Time & Related attacks & Tags & Type of false positive \\
\hline\hline
1 & 2015-12-28 12:19:05 PM  & 7 & P-101 & OP \\
~ & - 2015-12-28 12:20:46 PM & ~ & ~ \\
\hline
2 & 2015-12-29 02:32:36 PM & 17 & MV-101 & OP \\
~ & - 2015-12-29 02:36:34 PM & ~ & ~ \\
\hline
3 & 2015-12-29 02:39:42 PM & 17 & MV-101 & OP \\
~ & - 2015-12-29 02:43:39 PM & ~ & ~ \\
\hline
4 & 2015-12-30 01:51:41 PM & - & MV-101 & \textbf{TFP} \\
~ & - 2015-12-30 01:55:36 PM & ~ & ~ \\
\hline
5 & 2015-12-30 02:06:12 PM & - & MV-101 & \textbf{TFP} \\
~ & - 2015-12-30 02:09:55 PM & ~ & ~ \\
\hline
6 & 2015-12-30 06:20:01 PM & 26 & - & LT \\
~ & - 2015-12-30 06:21:07 PM & ~ & ~ \\
\hline
7 & 2016-01-01 10:58:18 AM & 32 & MV-101 & OP \\
~ & - 2016-01-01 11:02:11 AM & ~ & ~ \\
\hline
8 & 2016-01-01 11:10:06 AM & 32 & MV-101 & OP \\
~ & - 2016-01-01 11:13:38 AM & ~ & ~ \\
\hline
9 & 2016-01-01 03:02:23 PM & 33 & - & LT \\
~ & - 2016-01-01 03:04:04 PM & ~ & ~ \\
\hline
10 & 2016-01-02 11:32:16 AM & 38 & P-101 & OP \\
~ & - 2016-01-02 11:36:10 & ~ & ~ \\
\hline
11 & 2016-01-02 11:47:29 AM & 38 & P-101 & OP \\
~ & - 2016-01-02 11:51:04 AM & ~ & ~ \\
\hline
\end{tabular}
\end{center}
\label{tab:false_positives_p1}
\end{table}

\subsubsection{False positives in process 2}

The developed model for process 2 created eight false positives.
In Table~\ref{tab:false_positives_p2}, pairs (1, 2), (4, 5), and (6, 7) have the same related attacks.
Therefore, there were five independent false positives, and four of them were true.

Most false positives occurred at P-201, P-205, and MV-201.
P-201 is the NaCl injection jump; P-205 is the NaOCl injection pump.
In the training dataset, P-201 never changed.
We guess that the model regarded any change of P-201 as an attack.
AIT-201, a sensor for NaCl, caused a false positive (No. 8 in Table~\ref{tab:false_positives_p2}) because P-201 had changed the level of NaCl.

After the last false positive, the prediction errors of AIT-201 and AIT-203 went high.
AIT-201 and AIT-203 are sensors for NaCl and NaOCl, respectively.
We guess that the unexpected behaviors of P-201 and P-205 led to new but \textit{normal} dataset.

\begin{table}[t]
\caption{Analysis of false positives of process 2. OP indicates the number of false positives for attacks on other processes, and TFP indicates true false positives.}
\begin{center}
\begin{tabular}{c|c|c|c|c}
\hline
No. & Time & Related attacks & Tags & Type of false positive \\
\hline\hline
1 & 2015-12-29 07:18:41 PM & - & P-201, & \textbf{TFP} \\
~ & - 2015-12-29 07:22:14 PM & ~ & P-205  & ~ \\
\hline
2 & 2015-12-29 07:29:56 PM & - & P-201, & \textbf{TFP} \\
~ & - 2015-12-29 07:33:38 PM & ~ & P-205  & ~ \\
\hline
3 & 2015-12-29 09:23:58 PM & - & MV-201, & \textbf{TFP} \\
~ & - 2015-12-29 09:26:36 PM & ~ & P-201  & ~ \\
\hline
4 & 2015-12-29 09:44:08 PM & - & MV-201, & \textbf{TFP} \\
~ & - 2015-12-29 09:47:43 PM & ~ & P-201  & ~ \\
\hline
5 & 2015-12-29 09:52:23 PM & - & MV-201, & \textbf{TFP} \\
~ & - 2015-12-29 09:54:07 PM & ~ & P-201  & ~ \\
\hline
6 & 2015-12-29 11:12:18 PM & 16 & P-201, & OP \\
~ & - 2015-12-29 11:12:58 PM  & ~ & P-205  & ~ \\
\hline
7 & 2015-12-29 11:19:43 PM & 16 & P-201, & OP \\
~ & - 2015-12-29 11:20:37 PM & ~ & P-205 & ~ \\
\hline
8 & 2015-12-30 01:12:39 AM & - & P-201, & \textbf{TFP} \\
~ & - 2015-12-30 01:14:50 AM & ~ & AIT-201  & ~ \\
\hline
\end{tabular}
\end{center}
\label{tab:false_positives_p2}
\end{table}

\subsubsection{False positives in process 4}

In Table~\ref{tab:false_positives_p4}, there was one false positive in process 4, which came from attack 37 hitting process 5.

\begin{table}[t]
\caption{Analysis of false positives of process 4. OP indicates the number of false positives for attacks on other processes}
\begin{center}
\begin{tabular}{c|c|c|c|c}
\hline
No. & Time & Related attacks & Tags & Type of false positive \\
\hline\hline
1 & 2016-01-02 11:19:16 AM & 37 & - & OP \\
~ & - 2016-01-02 11:23:15 AM & ~ & ~ \\
\hline
\end{tabular}
\end{center}
\label{tab:false_positives_p4}
\end{table}

\section{Conclusion}
\label{sec:conclusion}

It is difficult to get internal specification and control logic of ICSs. If routine ICS operational data is the only information available, a data-driven approach is a proper way to develop security products.

We proposed an anomaly detection method for industrial control systems using sequence-to-sequence neural networks with attention.
Due to the difficulty of defining the abnormal state, the model learns the normal dataset in an unsupervised way.
In the detection phase, the model predicts future values based on the previously observed ones.
The difference between the model's prediction and the measured value is the key criterion to detect anomalies.

The alarm decision depends on the threshold, and heuristic hyper-parameters were necessary for this experiment.
Our proposed method is not dedicated to the dataset.
It can be generalized to train any ICS datasets and extract the decision grounds because the specification of operational data and control logic inside was not required.
It is also able to detect anomalies with only the dataset from normal operations.

\bibliographystyle{splncs04}
\bibliography{bib.bib}

\begin{thebibliography}{10}
\providecommand{\url}[1]{\texttt{#1}}
\providecommand{\urlprefix}{URL }
\providecommand{\doi}[1]{https://doi.org/#1}

\bibitem{ahmad2017anomaly}
Ahmad, S., Lavin, A., Purdy, S., Agha, Z.: Unsupervised real-time anomaly
  detection for streaming data. Neurocomputing  \textbf{262},  134 -- 147
  (2017)

\bibitem{attention}
Bahdanau, D., Cho, K., Bengio, Y.: Neural machine translation by jointly
  learning to align and translate. In: 3rd International Conference on Learning
  Representations, {ICLR} 2015, San Diego, CA, USA, May 7-9, 2015, Conference
  Track Proceedings (2015)

\bibitem{chen2018learning}
Chen, Y., Poskitt, C.M., Sun, J.: Learning from mutants: Using code mutation to
  learn and monitor invariants of a cyber-physical system. In: 2018 IEEE
  Symposium on Security and Privacy (SP). pp. 648--660. IEEE (2018)

\bibitem{choi2018dataset}
Choi, S., Yun, J.H., Kim, S.K.: A comparison of ics datasets for security
  research based on attack paths. In: The 13th International Conference on
  Critical Information Infrastructures Security (CRITIS) (2018)

\bibitem{formby2016s}
Formby, D., Srinivasan, P., Leonard, A., Rogers, J., Beyah, R.A.: Who's in
  control of your control system? device fingerprinting for cyber-physical
  systems. In: Network and Distributed Systems Security (NDSS) (2016)

\bibitem{goh2016dataset}
Goh, J., Adepu, S., Junejo, K.N., Mathur, A.: A dataset to support research in
  the design of secure water treatment systems. In: The 11th International
  Conference on Critical Information Infrastructures Security (CRITIS) (2016)

\bibitem{goh2017anomaly}
Goh, J., Adepu, S., Tan, M., Lee, Z.S.: Anomaly detection in cyber physical
  systems using recurrent neural networks. In: 2017 IEEE 18th International
  Symposium on High Assurance Systems Engineering (HASE). pp. 140--145. IEEE
  (2017)

\bibitem{lstm}
Hochreiter, S., Schmidhuber, J.: Long short-term memory. Neural Comput.
  \textbf{9}(8),  1735--1780 (Nov 1997)

\bibitem{adam}
Kingma, D.P., Ba, J.: Adam: {A} method for stochastic optimization. In: 3rd
  International Conference on Learning Representations, {ICLR} 2015, San Diego,
  CA, USA, May 7-9, 2015, Conference Track Proceedings (2015),
  \url{http://arxiv.org/abs/1412.6980}

\bibitem{israel}
Kravchik, M., Shabtai, A.: Detecting cyber attacks in industrial control
  systems using convolutional neural networks. In: Proceedings of the 2018
  Workshop on Cyber-Physical Systems Security and PrivaCy. pp. 72--83. CPS-SPC
  '18 (2018)

\bibitem{lemay2016providing}
Lemay, A., Fernandez, J.M.: Providing scada network data sets for intrusion
  detection research. In: Proceedings of the 9th USENIX Conference on Cyber
  Security Experimentation and Test. pp.~6--6. CSET'16, Berkeley, CA, USA
  (2016)

\bibitem{li2019mad}
Li, D., Chen, D., Shi, L., Jin, B., Goh, J., Ng, S.: {MAD-GAN:} multivariate
  anomaly detection for time series data with generative adversarial networks.
  CoRR  \textbf{abs/1901.04997} (2019)

\bibitem{lin2017timing}
Lin, C.Y., Nadjm-Tehrani, S., Asplund, M.: Timing-based anomaly detection in
  scada networks. In: International Conference on Critical Infrastructures
  Security (CRITIS) (2017)

\bibitem{mitchell2014survey}
Mitchell, R., Chen, I.R.: A survey of intrusion detection techniques for
  cyber-physical systems. ACM Computing Surveys (CSUR)  \textbf{46}(4), ~55
  (2014)

\bibitem{mitchell2013behavior}
Mitchell, R., Chen, R.: Behavior-rule based intrusion detection systems for
  safety critical smart grid applications. IEEE Transactions on Smart Grid
  \textbf{4}(3),  1254--1263 (2013)

\bibitem{amsgrad}
Reddi, S.J., Kale, S., Kumar, S.: On the convergence of adam and beyond. In:
  6th International Conference on Learning Representations, {ICLR} 2018,
  Vancouver, BC, Canada, April 30 - May 3, 2018, Conference Track Proceedings
  (2018)

\bibitem{rodofile2017process}
Rodofile, N.R., Schmidt, T., Sherry, S.T., Djamaludin, C., Radke, K., Foo, E.:
  {Process Control Cyber-Attacks and Labelled Datasets on S7Comm Critical
  Infrastructure}. In: Information Security and Privacy. pp. 452--459 (2017)

\bibitem{kaspersky}
Shalyga, D., Filonov, P., Lavrentyev, A.: Anomaly detection for water treatment
  system based on neural network with automatic architecture optimization.
  DISE1 Workshop, International Conference on Machine Learning (ICML) (2018)

\bibitem{seq2seq}
Sutskever, I., Vinyals, O., Le, Q.V.: Sequence to sequence learning with neural
  networks. In: Ghahramani, Z., Welling, M., Cortes, C., Lawrence, N.D.,
  Weinberger, K.Q. (eds.) Advances in Neural Information Processing Systems 27,
  pp. 3104--3112 (2014)

\bibitem{yun2018similarity}
Yun, J., Hwang, Y., Lee, W., Ahn, H., Kim, S.: Statistical similarity of
  critical infrastructure network traffic based on nearest neighbor distances.
  In: Research in Attacks, Intrusions, and Defenses - 21st International
  Symposium, {RAID} 2018. pp. 577--599 (2018)

\end{thebibliography}

\end{document}